\documentclass[iicol,sn-basic]{sn-jnl}

\usepackage{graphicx}%
\usepackage{multirow}%
\usepackage{amsmath,amssymb,amsfonts}%
\usepackage{amsthm}%
\usepackage{mathrsfs}%
\usepackage[title]{appendix}%
\usepackage{xcolor}%
\usepackage{textcomp}%
\usepackage{manyfoot}%
\usepackage{booktabs}%
\usepackage{algorithm}%
\usepackage{algorithmicx}%
\usepackage{algpseudocode}%
\usepackage{listings}%
\usepackage{amsmath}
\usepackage{float}

\begin{document}

\title[Article Title]{TCDiff\texttt{++}: An End-to-end Trajectory-Controllable Diffusion Model for Harmonious Music-Driven Group Choreography}


\author[1,2]{\fnm{Yuqin} \sur{Dai}}\email{daiy@njust.edu.cn}
\equalcont{Equal contribution}

\author[1]{\fnm{Wanlu} \sur{Zhu}}\email{wanluzhu@njust.edu.cn}
\equalcont{Equal contribution}

\author[2]{\fnm{Ronghui} \sur{Li}}\email{lrh22@mails.tsinghua.edu.cn}

\author[2]{\fnm{Xiu} \sur{Li}}\email{li.xiu@sz.tsinghua.edu.cn}

\author*[3]{\fnm{Zhenyu} \sur{Zhang}}\email{zhenyuzhang@nju.edu.cn}

\author*[1]{\fnm{Jun} \sur{Li}}\email{junli@njust.edu.cn}

\author*[1]{\fnm{Jian} \sur{Yang}}\email{csjyang@njust.edu.cn}

\affil[1]{PCA Lab, Key Lab of Intelligent Perception and Systems for High-Dimensional Information of Ministry of Education, School of Computer Science and Engineering, Nanjing University of Science and Technology, Nanjing, China}

\affil[2]{Shenzhen International Graduate School, Tsinghua University, Shenzhen, China}

\affil[3]{Nanjing University, Suzhou, China}

\abstract{
Music-driven dance generation has garnered significant attention due to its wide range of industrial applications, particularly in the creation of group choreography. During the group dance generation process, however, most existing methods still face three primary issues: \textit{\textbf{multi-dancer collisions}}, \textit{\textbf{single-dancer foot sliding}} and \textit{\textbf{abrupt swapping in the generation of long group dance}}. In this paper, we propose TCDiff\texttt{++}, a music-driven end-to-end framework designed to generate harmonious group dance. Specifically, to mitigate multi-dancer collisions, we utilize a dancer positioning embedding to encode temporal and identity information. Additionally, we incorporate a distance-consistency loss to ensure that inter-dancer distances remain within plausible ranges. To address the issue of single-dancer foot sliding, we introduce a swap mode embedding to indicate dancer swapping patterns and design a Footwork Adaptor to refine raw motion, thereby minimizing foot sliding. For long group dance generation, we present a long group diffusion sampling strategy that reduces abrupt position shifts by injecting positional information into the noisy input. Furthermore, we integrate a Sequence Decoder layer to enhance the model's ability to selectively process long sequences. 
Extensive experiments demonstrate that our TCDiff\texttt{++} achieves state-of-the-art performance, particularly in long-duration scenarios, ensuring high-quality and coherent group dance generation. 
\href{https://da1yuqin.github.io/TCDiffpp.website/}{Project Page}.
}




\maketitle

\section{Introduction}
\label{sec:intro} 
As one of the most expressive forms of art,  dance profoundly impacts cultural, cinematic, and academic domains \citep{artemyeva2018role, metaverse, dany, xue2024human}. This influence stems from the intricate process of choreography, where movement and music converge, synchronizing rhythm and structure to produce a harmonious and cohesive artistic expression. Traditionally, choreography has been a labor-intensive endeavor, spurring the development of automated models for dance creation. Consequently, music-driven choreography, initially centered on solo dancers \citep{aist++, edge, finedance}, has garnered considerable attention. With the growing demand for more immersive and interactive experiences, the focus has changed to multi-person choreography \citep{dany}. This evolution underscores the dual challenge of achieving synchronization in group movements while maintaining individuality within the ensemble, thereby enhancing the audience's engagement and creating a more dynamic experience \citep{schwartz1998passacaille}. Despite early recognition and exploration \citep{dany, aioz, gcd, codancers}, these approaches continue to face three significant problems:

\begin{figure*}[!t]
  \centering
  \includegraphics[width=0.97\linewidth]{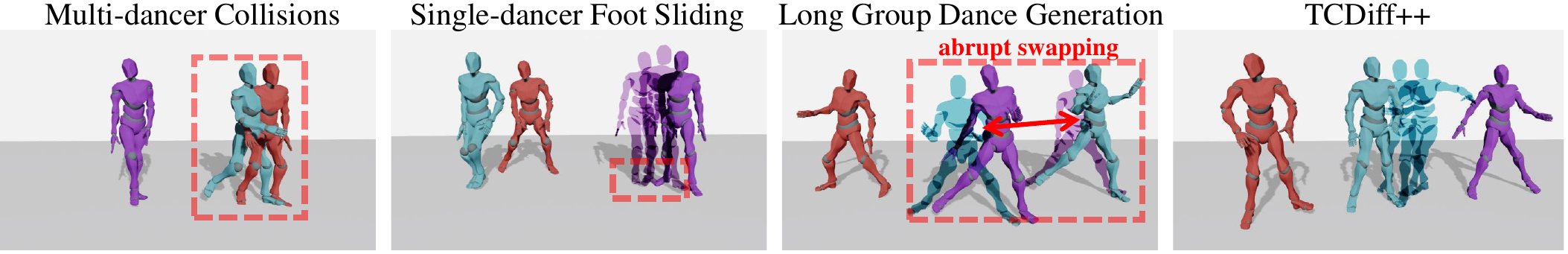}
  \vspace{-0.5mm}
  \caption{
  Visualizations of three key issues in baseline models: multi-dancer collisions \citep{tcdiff}, single-dancer foot sliding \citep{codancers}, and long group dance generation \citep{gcd}, where the blue man and the purple man suddenly swapped positions. In contrast, our approach eliminates these issues, delivering superior visual aesthetics.
  }
  \label{fig:intro_challenge}
\end{figure*}

\textbf{Multi-dancer collisions}.
Many group choreography frameworks \citep{aioz,gcd,dany} construct model inputs by concatenating the movements and coordinates of each dancer. However, this approach introduces a notable imbalance, as movements typically span over 100 dimensions, while coordinates are limited to just three. 
In group choreography, while dancers' positions can vary considerably, their movements often exhibit substantial similarity. For instance, more than 80\% of the movements in the AIOZ-GDance dataset \citep{aioz} are alike. This results in \textit{dancer ambiguity}, making it challenging for models to distinguish between individual dancers, which frequently leads to collisions, as illustrated in Figure \ref{fig:intro_challenge}.
In our previous work \citep{tcdiff}, we introduced the Dance-Trajectory Navigator (DTN), which focuses on key positional coordinates and uses a distance-consistency loss to prevent collisions. Although this approach reduces ambiguity, it still struggles to manage highly dynamic group interactions, particularly in scenarios involving numerous dancers and complex movements, often resulting in misalignment or collisions.

\textbf{Single-dancer foot sliding}. 
Foot sliding occurs when a dancer's feet appear to glide across the ground while the upper body maintains proper movement, as illustrated in Figure \ref{fig:intro_challenge}. This issue often arises from difficulties in accurately modeling the relationship between global trajectory and local body rotations \citep{longdancediff}. In multi-person choreography, dancer ambiguity makes this alignment even more difficult, further complicating footwork synchronization with movement.

\textbf{Long group dance generation}.
Existing models \citep{aioz, gcd, codancers, tcdiff} have demonstrated the ability to generate group dance movements lasting several seconds.  Unfortunately, a full group dance performance synchronized with music typically spans several minutes and, in the case of musicals, can extend to hours. 
This disparity highlights a critical gap in current capabilities, underscoring the urgent need for methods capable of generating long-duration group dance motions.
To tackle the challenge of long-duration generation, current methods first generate partially overlapping segments and then smooth the overlapping regions when stitching them together to form a coherent sequence. Consequently, ensuring consistency in these overlapping regions is essential for achieving seamless and continuous results. 

Drawing inspiration from the solo-dance model \citep{edge}, diffusion-based methods \citep{gcd, tcdiff} enforce this consistency during the sampling stage. However, group dance introduces an additional layer of complexity: it allows for \textit{\textbf{dancer swapping}}, where any two dancers can exchange positions while maintaining a reasonable formation. This flexibility introduces a contradiction when merging segments, as dancers' positions may vary across different sampling batches, resulting in positional inconsistencies. As a result, after stitching the segments, the same dancer may appear in discontinuous positions, causing abrupt and visually jarring shifts. These unnatural transitions significantly degrade the visual quality of long-duration group dance generation, posing a major challenge for achieving realistic and coherent performances.

A preliminary version of this work is the TCDiff (\textbf{T}rajectory-\textbf{C}ontrollable \textbf{Diff}usion)~\citep{tcdiff}, which was published in AAAI 2025. 
TCDiff proposes a two-stage framework that first uses a Dance-Trajectory Navigator (DTN) to predict dancers' coordinates, followed by a Trajectory-Specialist Diffusion (TSD) that generates movements based on these coordinates. 
The DTN focuses on learning coordinates with significant variations, while the TSD captures motion features with high similarity. This separation ensures distinct trajectory coordinates, effectively preventing dancer collisions. 

However, the decoupled nature of this two-stage process separates the modeling of footwork and movement, often resulting in disjointed actions and displacements. Additionally, the disregard for movement dynamics introduces uncertainties in trajectory generation. These uncertainties accumulate over time, particularly in long group dance sequences, leading to performance degradation such as sharp oscillations. As a result, TCDiff is ill-suited for long-duration scenarios. Furthermore, the model architecture lacks selectivity, which exacerbates its performance decline in extended sequences. The two-stage design also limits its potential to inform subsequent end-to-end models, thereby restricting TCDiff's broader applicability and impact.

To address the aforementioned issues, we upgrade the preliminary conference version~\citep{tcdiff} into a fully end-to-end model. This end-to-end design not only produces more coherent body movements and positions but also significantly enhances performance in long-duration scenarios. 
To be more specific,
\textit{\textbf{to mitigate multi-dancer collisions,}}
we design DPE, which enriches feature representations with temporal and dancer identity information to reduce ambiguity and collisions.
Additionally, while the distance-consistency loss was originally applied only within the DTN module, in this work, we apply it to the entire end-to-end model to enforce global consistency.
The end-to-end application of the distance-consistency loss leads to more effective control of the spacing between generated positions.
To further address dancer ambiguity, we implement a simple yet effective Fusion Projection module (FP) to enhance the distinction between dancers in high-dimensional space, effectively mitigating dancer ambiguity. 
\textit{\textbf{For single-dancer foot sliding}}, 
we newly integrate swap mode information, which specifies the start and end positions of dancers, thereby reducing spatial uncertainty caused by swap actions during generation. This enriched spatial information helps mitigate irregular foot sliding. Moreover, we refine the Footwork Adaptor plugin in TCDiff to directly adjust the generated results, improving the consistency between footwork and motion. 
\textit{\textbf{To facilitate long dance generation,}}
we remove the DTN module, which tends to cause oscillations due to spatial uncertainties in long-duration generation.
To further reduce abrupt position shifts caused by dancer swapping, we propose the long group diffusion sampling strategy. This method constrains the input noise and incorporates positional priors, improving position consistency and minimizing sudden swaps.
Additionally, we introduce the Sequence Decoder to strengthen the model's sequence selection capability, further enhancing its performance in long-duration generation scenarios. In summary, our main contributions are:
\begin{itemize}
\item To prevent multi-dancer collisions, we introduce a dancer positioning embedding to to encode temporal and dancer identity information and adapt the distance-consistency loss for end-to-end models, ensuring reasonable spacing. Additionally, a simple yet effective FP module is used to efficiently reduce dancer ambiguity.
\item For single-dancer foot sliding, we newly integrate swap mode information to reduce spatial uncertainty and enhance the Footwork Adaptor, improving footwork-motion consistency in the end-to-end model's outputs.
\item To facilitate long dance generation, our end-to-end architecture removes the DTN module to avoid oscillations from spatial uncertainties in long dance generation. We propose a long group diffusion sampling strategy, injecting positional information to reduce abrupt swapping, and introduce a Sequence Decoder layer to improve sequence selection for long-duration generation.
\item We propose TCDiff\texttt{++}, an end-to-end group dance generation model that surpasses the two-stage version in producing coherent body movements and handling long sequences. This widely applicable design also provides insights for future end-to-end models. Extensive experiments and analyses confirm the superiority of our approach over existing methods.
\end{itemize}

\section{Related Work}
\subsection{Single-dancer Generation}
\noindent Group dance generation extends single-dancer generation, but directly applying single-dancer methods to group dance presents significant challenges \citep{dancereview}. Early motion retrieval methods \citep{kovar2002pighin, fan2011example, ofli2011learn2dance, lee2013music} often produce deformed actions. Recent approaches use large datasets \citep{lee2019dancing, dancerevolution, valle2021transflower, aist++, finedance, enchantdance} and deep learning techniques, such as auto-regressive models \citep{alemi2017groovenet, yalta2019weakly, ahn2020generative, dancecontrol, bailando} and generative models \citep{mnet, edge, lodge, li2024lodge++, zhou2023learning} , to generate motions. Recently, diffusion-based models \citep{diffusion, sohl2015deep, edge, ren2025realistic, guo2025controllable, yang2025unimumo} have emerged, achieving state-of-the-art performance with high diversity and fidelity.
Solo dance generation emphasizes realism, where artifacts like foot sliding are unacceptable \citep{longdancediff}. To address this, existing methods impose physical constraints through loss functions and foot contact labels \citep{zhang2021learning, zhang2023real, edge, lodge}.

Since music often lasts for several minutes, and some musicals extend for hours, improving performance in long-duration scenarios is crucial. To address long-duration generation, some works~\citep{zhang2024large, lodge, li2025melodance} adopt a progressive generation strategy, first producing choreography-rich dance segments and then generating transitions to seamlessly concatenate them into full-length sequences. 
However, single-dancer models often suffer from dancer ambiguity due to highly similar representations across dancers, leading to collisions in group choreography. To mitigate this, we propose a dancer positioning embedding (DPE) that encodes temporal and identity cues to distinguish dancers. In addition, we introduce a distance-consistency loss to keep inter-dancer spacing within reasonable limits.

\subsection{Multi-dancer Generation}
\noindent  Multi-dancer generation is an emerging field in its early stages.
To our knowledge, few studies \citep{dany, aioz, gcd, codancers, pdvae, tcdiff} have focused on scenarios with more than two dancers. 
Among them, GDanceR \citep{aioz} and GCD \citep{gcd} do not employ any specific structures to address motion representation imbalance, leading to dancer ambiguity.
CoDancers~\citep{codancers} splits group motions into single-dancer motions to achieve scalability by incrementally adding dancers to the formation. As a result, this division causes the models to acquire only local information, leading to suboptimal group visual effects.
TCDiff~\citep{tcdiff} is our previous version, a two-stage framework that first predicts dancers' coordinates and then generates corresponding movements based on obtained coordinates. This separation preserves global formation features and ensures distinct trajectory coordinates, thus preventing dancer collisions caused by ambiguity. 
However, the two-stage generation process separates footwork from movement modeling, often leading to disjointed actions and unnatural displacements. Therefore, there is an urgent need for a model architecture that can reduce dancer ambiguity while jointly capturing both global and local information. In this paper, we upgrade the previous TCDiff into an end-to-end framework capable of learning global and local features simultaneously. To further mitigate dancer ambiguity, we modify the original components and introduce new modules that impose additional spatial constraints.

\subsection{Diffusion Model}
\noindent In recent years, Diffusion Models (DM)~\citep{ho2020denoising,song2020denoising} have demonstrated outstanding performance in the content generation domain~\citep{bar2024lumiere,wu2024direct3d,xiong2024novel,li2024tp2o, li2024dispose,zhu2024champ}. 
Compared to earlier generative models like VAEs~\citep{kingma2013auto} and GANs~\citep{goodfellow2020generative}, diffusion models (DMs) offer clear advantages. VAEs often struggle with capturing complex, high-dimensional distributions, while GANs face instability and mode collapse due to their adversarial setup~\citep{xue2024human}. In contrast, DMs use probabilistic inference without adversarial training, leading to higher fidelity and more stable learning. Like previous methods, DMs also support conditional generation~\citep{bian2025motioncraft,yang2025unimumo}.
This ability is particularly important in motion generation, especially for human motion and dance synthesis~\citep{liang2024intergen, dabral2023mofusion}. 
In these tasks, conditional inputs, such as text~\citep{wang2023intercontrol,zhou2023ude,zhou2024learning} and audio~\citep{alexanderson2023listen,tseng2023edge}, guide the model to generate desired motions.
However, dancer swapping causes inconsistencies in overlapping segments during diffusion sampling, resulting in abrupt transitions. To mitigate this, we introduce a long group diffusion sampling strategy that constrains noise and leverages positional priors to ensure consistent, long-duration generation.

\begin{figure*}[!t]
  \centering
  \includegraphics[width=1\linewidth]{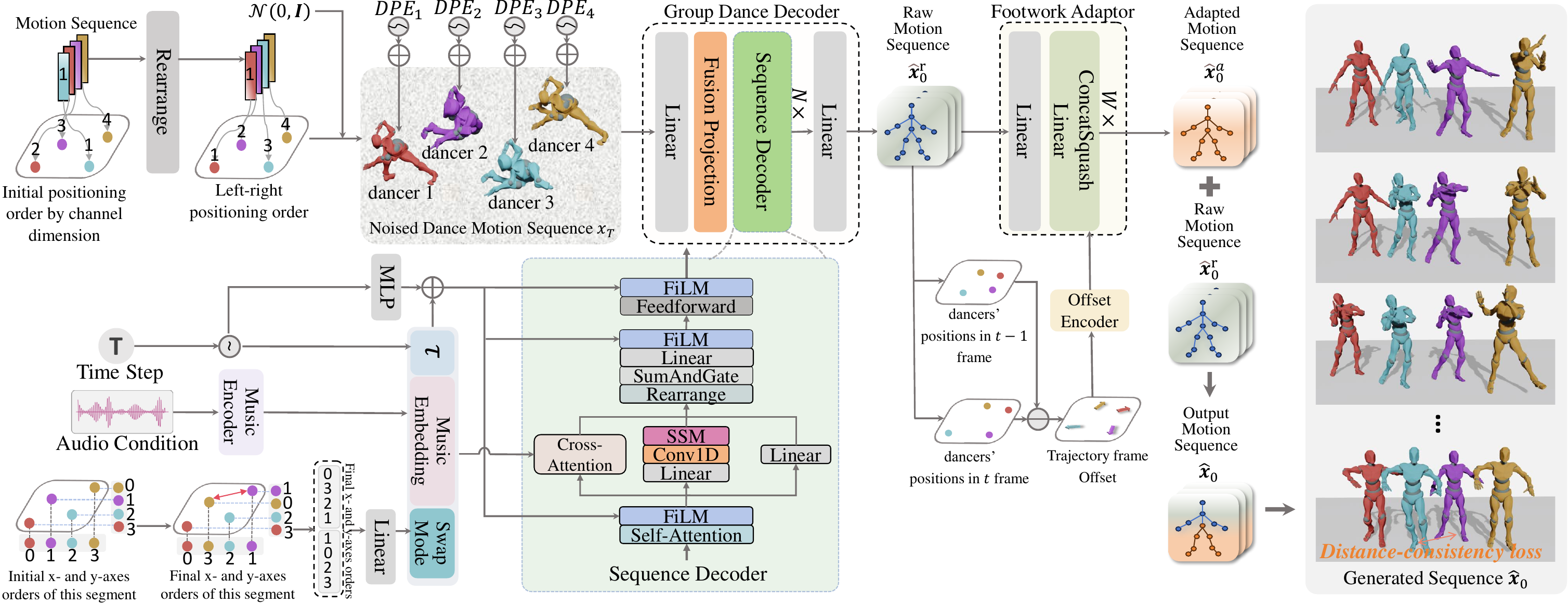}
  \vspace{-10pt}
  \caption{
  Our end-to-end TCDiff\texttt{++} framework comprises two key components: the Group Dance Decoder (GDD) and the Footwork Adaptor (FA). The GDD initially generates a raw motion sequence $\hat{\boldsymbol{x}}^r$ without trajectory overlap based on the given music. Subsequently, the FA refines the foot movements by leveraging the positional information of the raw motion, producing an adapted motion $\hat{x}_0^a$ with improved footstep actions to reduce foot sliding. Finally, the adapted footstep movements are incorporated into the raw motion, yielding a harmonious dance sequence $\hat{\boldsymbol{x}}_0$ with stable footwork and less dancer collisions. 
  Compared to the previous two-stage version, TCDiff\texttt{++} requires only a single training stage, demonstrating better footwork-motion coherence performance. 
  } 
  \label{fig:tcd_framwork}
\end{figure*}

\section{Background}
\label{sec:Background}
\subsection{Problem Definition}
\label{sec:ProblemDefinition}
\noindent Given an input music sequence $\mathcal{M}=\left\{\boldsymbol{m}_i\right\}_{i=1}^L$ with $i=\left\{1,\dots,L\right\}$ indicates the index of the music frames, group choreography is to generate a corresponding group dance movement sequence $\boldsymbol{x}=\left\{\boldsymbol{x}^{(i)}\right\}_{i=1}^L$. Here, $\boldsymbol{x}^{(i)}=\left\{\boldsymbol{x}^{(i),c}\right\}_{c=1}^C$, where $\boldsymbol{x}^{(i),c}$ is the generated pose of $c$-th dancer at $i$-frame. $\boldsymbol{x}^c=\left\{\boldsymbol{x}^{(i),c}\right\}_{i=1}^L$ represents the movement sequence of the $c$-th dancer. For simplicity, we use $\boldsymbol{x}_t$ to represent $\{\boldsymbol{x}_{t}^{(i),c}\}$ at each timestep $t$ of the diffusion process.

\subsection{Motion and Music Features} 
The dancer's motion is modeled using a 24-joint SMPL ~\citep{smpl}, with each joint's pose $\boldsymbol{d} \in \mathbb{R}^{24 \times 6=144}$ expressed in 6D rotation \citep{rot6d}. This model also includes binary contact indicators $\boldsymbol{f} \in \mathbb{R}^{4}$ for the heels and toes, and a 3D root position $\boldsymbol{p} \in \mathbb{R}^{3}$, forming a complete motion descriptor $\boldsymbol{x}=[\boldsymbol{f},\boldsymbol{p},\boldsymbol{d}] \in \mathbb{R}^{144+4+3=151}$. 
For audio, we use the Librosa \citep{librosa} to build a music encoder that decomposes each audio clip into harmonic and percussive components and extracts mel-spectrograms, MFCCs (and deltas), chroma, onset/beat indicators, and a tempogram. The resulting features are concatenated and cached as fixed conditioning inputs.

\subsection{Diffusion for Dance Generation}
\noindent We utilize a diffusion framework \citep{diffusion} to obtain dance sequence. The method constructs a Markov chain that progressively perturbs the original data $\boldsymbol{x}_0$ toward a standard Gaussian distribution $\boldsymbol{x}_T \sim \mathcal{N}\left(0, \boldsymbol{I}\right)$ through $T$ iterative noise injection stages. This forward process is defined as:
\begin{equation}
q\left(\boldsymbol{x}_t \mid \boldsymbol{x}_{t-1}\right)=\mathcal{N}\left(\sqrt{\alpha_t} \boldsymbol{x}_{t-1},\left(1-\alpha_t\right) \boldsymbol{I}\right),
\end{equation}
where $\alpha_t \in \left(0,1\right)$ denotes a predefined decay coefficient sequence, and $\boldsymbol{I}$ is the identity matrix. The reverse process aims to denoise $\boldsymbol{x}_T$ back to $\boldsymbol{x}_0$. Since estimating $q\left(\boldsymbol{x}_{t-1}\mid \boldsymbol{x}_t\right)$ is challenging, the network $p_{\theta}$ is trained to approximate $\boldsymbol{x}_0$: 
\begin{equation}
p_{\theta}\left(\boldsymbol{x}_{t-1} \mid \boldsymbol{x}_t\right) = \mathcal{N}\left(\boldsymbol{x}_{t-1}; \mu_{\theta}\left(\boldsymbol{x}_t, t\right), \Sigma_{\theta}\left(\boldsymbol{x}_t, t\right)\right).
\end{equation}
To synthesize group dance motion sequences based on music rhythm, we extend the conditional diffusion framework. Given the music condition $\mathcal{M}$ \citep{edge} and the dancer spatial condition swap mode $S$, the generation process reverses the forward diffusion trajectory through a learned denoising function $D\left(\boldsymbol{x}_t, t, \mathcal{M}, S\right)$, which estimates the clean motion $\boldsymbol{x}$ at each time step $t$.

\subsection{SSM Backbone}
State Space Models (SSMs)~\citep{gu2022efficiently,mamba,mamba2}, have shown strong performance in modeling long sequences. Their effectiveness comes from three key features: selective control over internal state transitions, efficient linear-time computation, and the ability to capture global context across the entire sequence~\citep{MambaLLIE}.
At their core, SSMs transform an input signal \( x(t) \in \mathbb{R} \) into an output \( y(t) \in \mathbb{R} \) through a hidden state \( h(t) \in \mathbb{R}^N \), governed by a linear ordinary differential equation (ODE):
\begin{equation}
h'(t) = \mathbf{A} h(t) + \mathbf{B} x(t), \quad
y(t) = \mathbf{C} h(t),
\end{equation}
where \( N \) is the hidden state size, \( \mathbf{A} \in \mathbb{R}^{N \times N} \) controls the state evolution, and \( \mathbf{B} \in \mathbb{R}^{N \times 1} \), \( \mathbf{C} \in \mathbb{R}^{1 \times N} \) are input and output projection matrices.
To support discrete-time computation, Mamba~\citep{mamba} discretizes the system using zero-order hold (ZOH), resulting in:
{\small
\begin{align}
\overline{\mathbf{A}} &= \exp(\Delta \mathbf{A}) \label{eq:abar}, \\
\overline{\mathbf{B}} &= (\Delta \mathbf{A})^{-1} (\exp(\Delta \mathbf{A}) - \mathbb{I}) \Delta \mathbf{B}, \label{eq:bbar}
\end{align}
}
where \( \Delta \) is the input-dependent step size that controls temporal granularity: small \( \Delta \) captures fine-grained dynamics (e.g., rapid motions), while large \( \Delta \) captures long-term patterns.

The discretized model computes outputs through a global convolution with a structured kernel \( \overline{\mathbf{K}} \):
{\small
\begin{equation}
\begin{aligned}
\overline{\mathbf{K}} &= (\mathbf{C} \overline{\mathbf{B}}, \mathbf{C} \overline{\mathbf{A} \mathbf{B}}, \dots, \mathbf{C} \overline{\mathbf{A}}^{L-1} \overline{\mathbf{B}}), \\
\mathbf{y} &= \mathbf{x} * \overline{\mathbf{K}},
\end{aligned}
\end{equation}
}
where \( L \) is the input sequence length. This architecture allows the model to efficiently incorporate information from the entire sequence, improving its ability to handle long-range dependencies.

\section{Methodology}
\label{sec:method}
In this section, we introduce TCDiff\texttt{++}, an end-to-end framework that can generate harmonious group dance movements according to the input music, as shown in Figure \ref{fig:tcd_framwork}. 
Compared to the previous two-stage version, TCDiff\texttt{++} requires only a single training stage to achieve harmonious dance generation, free from the spatial uncertainty caused by the process of generating motions and trajectories separately in the previous version.
As a result, it demonstrates better footwork-motion coherence performance. 
TCDiff\texttt{++} consists of a Group Dance Decoder (GDD) and a Footwork Adaptor (FA). The GDD first generates a raw motion sequence $\hat{\boldsymbol{x}}^r$ without trajectory overlap based on the music input. Then, the FA further modulates the foot movements using the raw motion’s positional information, generating adapted motion $\hat{\boldsymbol{x}}^a$ with footstep actions that mitigate foot sliding. Finally, we integrate the footstep movements into the raw motion, resulting in a harmonious dance sequence $\hat{\boldsymbol{x}}_0$ with grounded footwork and no dancer collisions. 

\subsection{Group Dance Decoder}
\label{Sec:GDD}
The Group Dance Decoder (GDD) reconstructs the input noisy dance motion sequence \( \boldsymbol{x}_T \) into a clean raw motion sequence \( \hat{\boldsymbol{x}}^r \). GDD uses conditional generation to obtain \( \hat{\boldsymbol{x}}^r \), incorporating not only the music condition \( \mathcal{M} \), but also the diffusion time step \( T \) and the dancer spatial condition swap mode \( S \) to guide the generation of dancer motions.
Namely, the generation of GDD, denoted as $D\left(\cdot\right)$, executes a reverse diffusion process by estimating $D\left(\boldsymbol{x}_T, T, \mathcal{M}, S\right) \approx \boldsymbol{\hat{x}}^r$.
To be more specific, GDD first incorporates the Dancer Positioning Embedding (DPE) into the input features to enhance spatial representation. The Fusion Projection then processes the representation to reduce dancer ambiguity. Next, the Sequence Decoder enhances feature extraction through sequence selection, producing the final raw motion $\boldsymbol{\hat{x}}^r$.
We provide a detailed introduction to the modules of GDD in the following sections, starting with the Dancer Positioning Embedding (DPE).

\textbf{Dancer Positioning Embedding (DPE) }
encodes temporal and identity information to uniquely represent each dancer across time.
We first rearrange the clean ground-truth motion sequences such that dancers are ordered from left to right according to their horizontal (x-axis) positions in the initial frame, ensuring a consistent spatial layout across samples.
After sorting, we apply the diffusion forward process to the rearranged motion sequences, injecting Gaussian noise, and obtain the noisy data $\boldsymbol{x}_{T}$.
To encode temporal and dancer identity information into the noisy data $\boldsymbol{x}_{T}$, we first compute the DPE using the sinusoidal positional encoding from Transformer~\citep{attention}. The positional encoding for any index $u$ (which can be either a time index $i$ or a dancer index $c$) at embedding dimension $2k$ or $2k+1$ is computed as:
\begin{equation}
    \mathrm{PE}(u, 2k) = \sin\left(\frac{u}{10000^{\frac{2k}{d}}}\right),
\end{equation}
\begin{equation}
    \mathrm{PE}(u, 2k+1) = \cos\left(\frac{u}{10000^{\frac{2k}{d}}}\right),
\end{equation}
where $k = 0, 1, \ldots, \frac{d}{2}-1$, and $d$ is the embedding dimension.
For each dancer $c$ at frame $i$, the DPE vector is defined as:
\begin{equation}
    \boldsymbol{e}_{c,i} = \mathrm{PE}(i) + \mathrm{PE}(c).
\end{equation}

We integrate the DPE into the noisy motion data by performing an element-wise addition for each pose vector:
\begin{equation}
    \boldsymbol{x}_{T}^{\prime\,(i), c} = \boldsymbol{x}_{T}^{(i), c} + \boldsymbol{e}_{c,i}.
\end{equation}

This yields the final encoded noisy sequence:
{\small 
\begin{equation}
    \boldsymbol{x}'_{t} = \left\{ \boldsymbol{x}_{t}^{\prime\,(i), c} \;\middle|\; i = 1,\dots, L; \; c = 1,\dots, C \right\}.
\end{equation}
}

This ensures that even after noise injection, the model retains both dancer-specific and temporal information, enhancing the consistency and controllability of generated motion sequences.

\begin{figure*}[!t]
  \centering
  \includegraphics[width=0.75\linewidth]{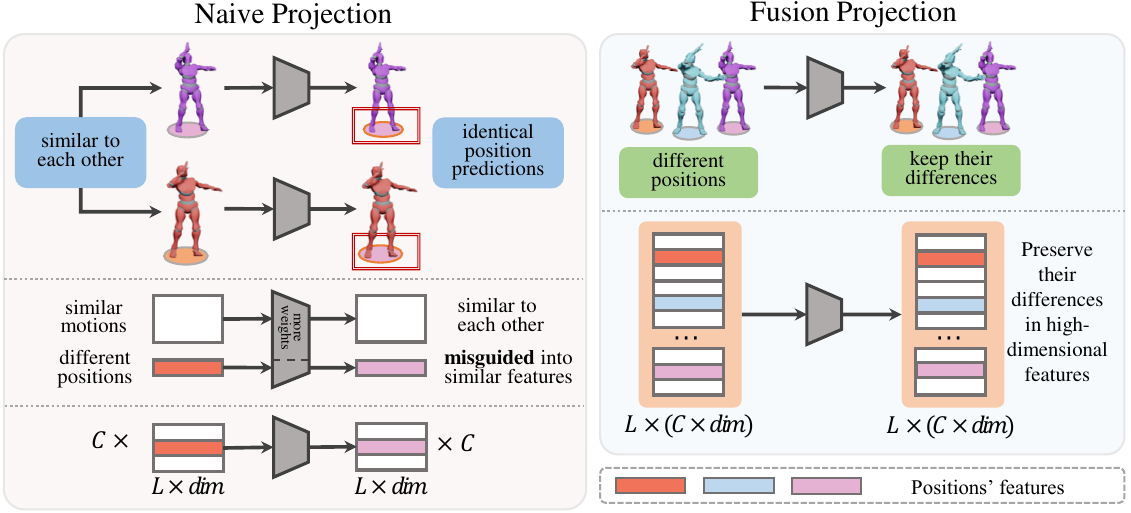}
  \caption{Our Fusion Projection (FP) module addresses the issue of dancer ambiguity. Imbalanced feature representations can cause positions to be misinterpreted as similar, leading to identical predictions. The FP module increases input dimensionality to enhance dancer differentiation, preserving positional differences and reducing collisions.
  }
  \label{fig:fp}
\end{figure*}

\textbf{Fusion Projection (FP).}
\label{sec:FusionProjection}
FP increases the features to a high-dimensional space to amplify differences between dancers, thereby reducing dancer ambiguity.
The FP module is designed under the principle that data in high-dimensional feature spaces can be more easily differentiated. To amplify the differences between dancers in this space, the FP module increases the input dimensionality by stacking the features of all dancers, as shown in Figure~\ref{fig:fp}. 
To be more specific, given an input feature tensor $\boldsymbol{x} \in \mathbb{R}^{C\times L\times d}$ with $C$ channels (also the number of dancers), $L$ frames, and hidden dimension $d$. We first concatenate the channel-wise features across all frames to form a reshaped tensor $\boldsymbol{x}' \in \mathbb{R}^{L\times \left(C \times d \right)}$.
These high-dimensional features are then mapped through a Multi-Layer Perceptron (MLP), producing latent space feature vectors for all dancers. Dancer-specific features are obtained by rearranging these vectors.
Unlike naive projection that maps a single dancer's features with an MLP, this approach maps the features of all dancers simultaneously, using a higher-dimensional space. 
This enhances the model's ability to capture differences at the group level in a higher-dimensional space, rather than at the individual level, effectively alleviating dancer ambiguity. Evidence for this is provided in the ablation study.

\textbf{Sequence Decoder (SD).}
The SD enhances the model's ability to selectively focus on long-duration information, utilizing the SSM architecture to improve selection in sequential data.
Specifically, the motion feature is first processed through self-attention for localized fine-grained modeling, followed by the SSM~\citep{mamba,mamba2} for feature selection.
We employ Cross-Attention to incorporate conditioning into the motion feature processing, while timestep information $T$, music conditioning $\mathcal{M}$, and dancer spatial condition swap mode $S$ are concatenated and injected into Feature-wise Linear Modulation (FiLM) \citep{perez2018film} to enhance the incorporation of conditional information.

\textbf{Swap Mode (SM)} identifies whether dancers change their relative positions between the start and end of each segment. Taking a 4-dancer group performance as an example, as illustrated in Fig.~\ref{fig:tcd_framwork}, we assign IDs 0, 1, 2, and 3 to the four dancers based on their initial positions along the x- and y-axes, ordered from left to right and top to bottom. Consequently, the initial x- and y-axis orders are always \([0, 1, 2, 3]\). At the final frame, the new x- and y-axis orders are recorded and concatenated into a single index sequence, such as \([0, 3, 2, 1, 1, 0, 2, 3]\). As shown in the figure, this index sequence clearly captures the position swap between the two dancers (indicated by the red arrow), effectively encoding the swap information. This sequence is subsequently embedded into a high dimension swap mode vector $\boldsymbol{S} \in \mathbb{R}^{d}$, which is then integrated with other conditioning inputs, including the music and the diffusion timestep.

\begin{figure*}[!t]
  \centering
  \includegraphics[width=0.97\linewidth]{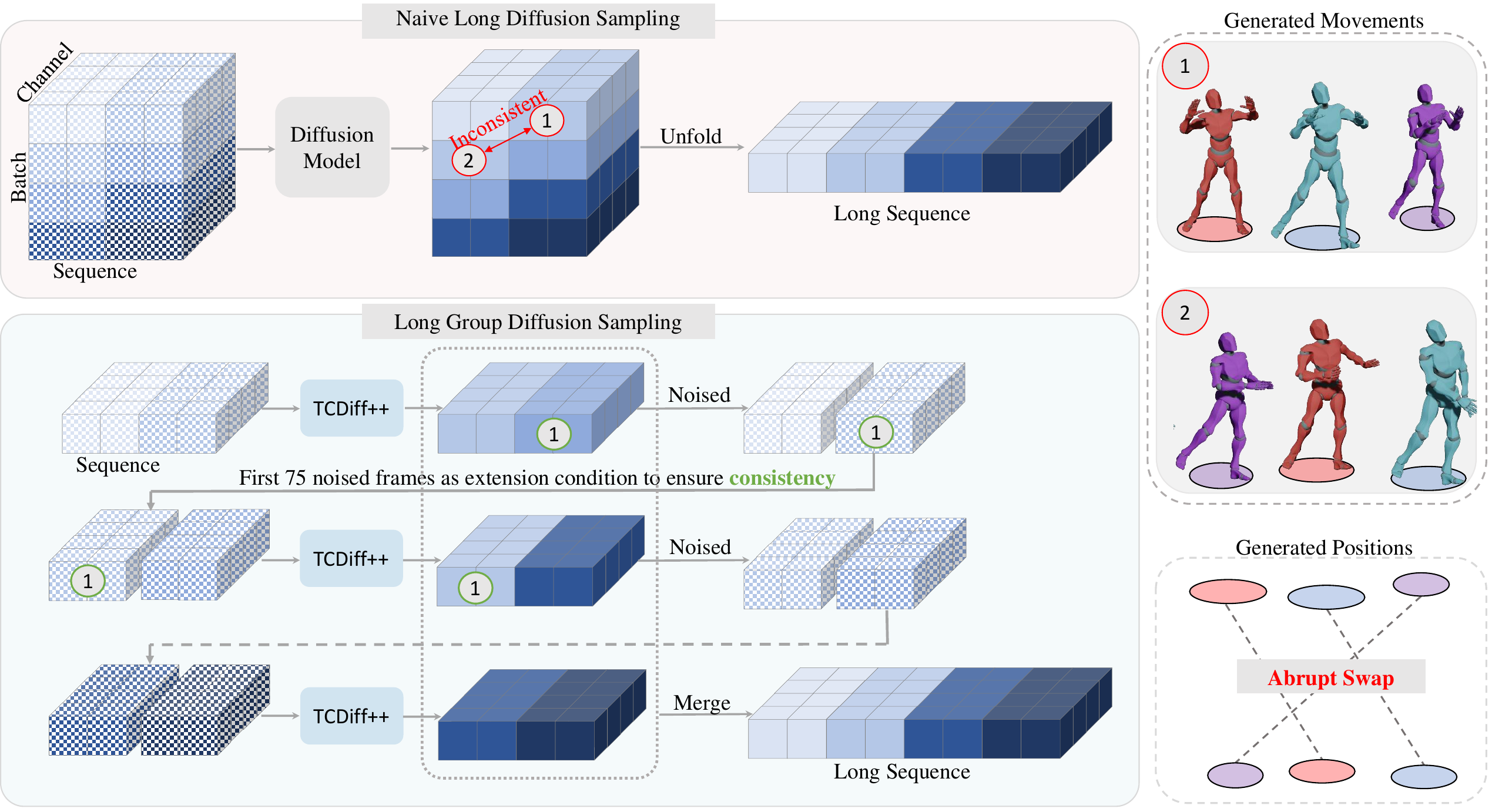}
  \caption{
  Our Long Group Diffusion Sampling (LGDS) method initially generates segments with partial overlap, which are then merged to form a complete sequence. Unlike naive sampling, LGDS enforces consistency during the input phase rather than the sampling phase. This approach reduces randomness and ensures cleaner positional information during generation, thereby reducing abrupt swap.
  }
  \label{fig:longsampling}
\end{figure*}

\subsection{Footwork Adaptor}
\label{sec:FA}
Since positional changes are mainly caused by footwork, our Footwork Adaptor (FA) module exclusively adjusts the dancers' lower body movements rather than the whole body motion.
Unlike the previous two-stage version, our end-to-end TCDiff\texttt{++} does not have access to clean dancer trajectory coordinates in advance for conditional generation and trajectory modulation. To address this, we use the raw motion $\hat{\boldsymbol{x}}^r$ generated by GDD as a substitute, extracting its trajectory coordinates as conditions. We then employ FA to refine footwork, obtaining the adapted motion $\hat{\boldsymbol{x}}^a$, thereby reducing foot sliding occurrences.
To be specific, the raw motion sequence generated by GDD, $\hat{\boldsymbol{{x}}}^r = [\boldsymbol{f}^r, \boldsymbol{p}^r, \boldsymbol{d}^r]$, includes the dancer's position data $\boldsymbol{p}^r$. We extract the coordinates and compute the frame-wise velocity by subtracting adjacent frames:
\begin{equation}
\mathcal{V}=\{\boldsymbol{p}_i^r-\boldsymbol{p}_{i-1}^r\}_1^L.
\end{equation}
This velocity conditions the Footwork Adaptor module $FA(\cdot)$ for correction, ultimately yielding the adapted motion $\hat{\boldsymbol{x}}_0^a$, as formulated below:
\begin{equation}
\hat{\boldsymbol{x}}^a=FA(\hat{\boldsymbol{x}}^r, \mathcal{V}).
\end{equation}
Specifically, the Footwork Adaptor $FA(\cdot)$ consists of a linear layer and a ConcatSquashLinear layer, as shown in Figure \ref{fig:tcd_framwork}. This design has been proven effective in various coordinate prediction domains \citep{mid, 3dpointdiffusion}.
Since positional changes are predominantly driven by footwork, we replace the footwork-related motions to the adopted version to obtain the final output. Specifically, we use the footwork-related motions from the adapted motion $\hat{\boldsymbol{x}}^a$ while preserving the expressive upper-body actions from the raw motion $\hat{\boldsymbol{x}}^r$, yielding the final generated result.
Therefore, we split $\hat{\boldsymbol{x}}^r$ and $\hat{\boldsymbol{x}}^a$ into upper and lower body components. 
In other words, 
$\hat{\boldsymbol{x}}^r = \{\hat{\boldsymbol{x}}^r_{\text{upper}}, \hat{\boldsymbol{x}}^r_{\text{lower}}\}$ 
and 
$\hat{\boldsymbol{x}}^a = \{\hat{\boldsymbol{x}}^a_{\text{upper}}, \hat{\boldsymbol{x}}^a_{\text{lower}}\}$. 

We then take the upper-body part from $\hat{\boldsymbol{x}}^r$ and the lower-body part from $\hat{\boldsymbol{x}}^a$, which contains the detailed footwork, to form the final output:
\begin{equation}
\widehat{\boldsymbol{x}}_0 = \hat{\boldsymbol{x}}^r_{\text{upper}} \oplus \hat{\boldsymbol{x}}^a_{\text{lower}}
\end{equation}
This approach allows the FA module to effectively combine footwork with positional information, thus refining the footwork in the raw sequence.
We show our effectiveness in Section \ref{sec:faab}. 

\subsection{Training Loss}
\label{sec:losses}
Our TCDiff\texttt{++} integrates the losses from the previous two-stage framework into a single framework, enabling more efficient end-to-end training. Specifically, the simple loss $\mathcal{L}_\text{simple}$ of the diffusion model is used to ensure that the reconstructed results are close to the ground truth:
\begin{equation}
    \mathcal{L}_{\text{simple}} = \mathbb{E}_{x,t} \left[ \| \boldsymbol{x}_0 - D(\boldsymbol{x}_T, T, \mathcal{M}, S) \|_2^2 \right],
\end{equation}
We adopt the joint velocity loss $\mathcal{L}_{\text{vel}}$ to ensure motion continuity between frames, the foot contact loss $\mathcal{L}_{\text{con}}$ to further constrain foot-ground contact, and the Forward-Kinematic loss $\mathcal{L}_{\text{FK}}$ to maintain the spatial consistency of the dancer's skeletons:
\begin{footnotesize}
\begin{equation}
\mathcal{L}_{\text {FK }} = \frac{1}{N} \sum_{i=1}^N \left\| F K\left( \boldsymbol{x}_0^{(i)} \right) - F K\left( \hat{\boldsymbol{x}}_0^{(i)} \right) \right\|_2^2,
\end{equation}
\begin{equation}
\mathcal{L}_{\text {vel }} = \frac{1}{N-1} \sum_{i=1}^{N-1} \left\| \left( \boldsymbol{x}^{(i+1)}_0 - \boldsymbol{x}^{(i)}_0 \right) - \left( \hat{\boldsymbol{x}}^{(i+1)}_0 - \hat{\boldsymbol{x}}^{(i)}_0 \right) \right\|_2^2,
\end{equation}
\begin{equation}
\mathcal{L}_{\text {con }} = \frac{1}{N-1} \sum_{i=1}^{N-1} \left\| \left( F K\left( \hat{\boldsymbol{x}}^{(i+1)}_0 \right) - F K\left( \hat{\boldsymbol{x}}^{(i)}_0 \right) \right) \cdot \hat{\boldsymbol{d}}^{(i)} \right\|_2^2.
\end{equation}
\end{footnotesize}
Here, $FK(\cdot)$ is the forward kinematic function that calculates the positions of joints given the 6D rotation motion.
Further, we integrate our distance-consistency loss $\mathcal{L}_{D}$:
\begin{footnotesize}
\begin{align}
\Delta\boldsymbol{p}^{(w),ij}= \left(\boldsymbol{p}^{(w),i}-\boldsymbol{p}^{(w),j}\right)-\left(\widehat{\boldsymbol{p}}^{(w),i}-\widehat{\boldsymbol{p}}^{(w),j}\right), \\
\mathcal{L}_{D}=\frac{1}{C-1} \sum_{w=1}^L\binom{C}{2}_{ij}\left\|\Delta\boldsymbol{p}^{(w),ij}\right\|_2^2,
\end{align}
\end{footnotesize}

\noindent which ensures that the spacing among dancers is within an appropriate range.
The overall objective of our proposed TCDiff\texttt{++} is:
\begin{align}
\mathcal{L}_{\mathrm{TCDiff\texttt{++}}} &= \lambda_{\text {sim}}\mathcal{L}_{\text{simple}} + \lambda_{\text{FK}} \mathcal{L}_{\text{FK}} + \lambda_{\text{vel}} \mathcal{L}_{\text{vel}} \notag \\
&\quad + \lambda_{\text{con}} \mathcal{L}_{\text{con}} + \lambda_{\text{D}} \mathcal{L}_{\text{D}}
\end{align}
where $\lambda_{\text {sim}} $, $\lambda_{\text {FK}} $, $\lambda_{\text {vel}} $, $\lambda_{\text {con}} $, and $\lambda_{\text {D}}$ are the balanced hyper-parameters.

\subsection{Long Group Diffusion Sampling}
\label{sec:longsampling}
We initially generate segments with partial overlap and subsequently merge them to construct a complete sequence.
Due to dancer swapping in group dance, dancers' positions can vary across different sampling epochs.
This makes it difficult to ensure consistency during the sampling stage, resulting in inconsistent positions during long-term generation.
Based on this finding, we propose Long Group Diffusion Sampling (LGDS) strategy to eliminate the uncertainty during sampling, resulting in more consistent generation.
LGDS can autoregressively extend short sequences by using the generated results as conditions, as illustrated in Figure~\ref{fig:longsampling}. This constrains the generation consistency during the input phase, rather than the sampling phase.
To be more specific, for a sequence of 150 frames, Instead of directly inputting random noise, we first sample a short clean segment $\widehat{\boldsymbol{{x}}}_{0}^{[0:150]}=\left\{ \widehat{\boldsymbol{{x}}}_{t}^{(1)}, \dots, \widehat{\boldsymbol{{x}}}_{t}^{(75)}, \dots, \widehat{\boldsymbol{{x}}}_{t}^{(150)}\right\}$, then add noise to obtain noised data $\widehat{\boldsymbol{{x}}}_T^{[0:150]}$. 
This noise data is used as input for the next step, recursively extending the sequence. 
The input can be represented as 
\begin{equation}
    \widehat{\boldsymbol{z}}_{T}^{[75:225]} = [ \widehat{\boldsymbol{x}}_{T}^{[75:150]}, \, \boldsymbol{z}_{T}^{[151:225]} ],
\end{equation}
where $[\cdot,\cdot]$ denotes the concatenation operation, $\boldsymbol{{z}}_{T}^{[151:225]}$ is a random noise, and $\widehat{\boldsymbol{{z}}}_{T}^{[75:225]}$ is the final input noise.
Compared to the existing method, this approach, which constrains the input noise, introduces less randomness and provides cleaner positional constraints during generation, thus effectively improving the consistency of the transitions in long-term generation.
We provide further analysis in the ablation study section, demonstrating the effectiveness of our approach.

\begin{figure*}[!t]
  \centering
  \includegraphics[width=0.98\linewidth]{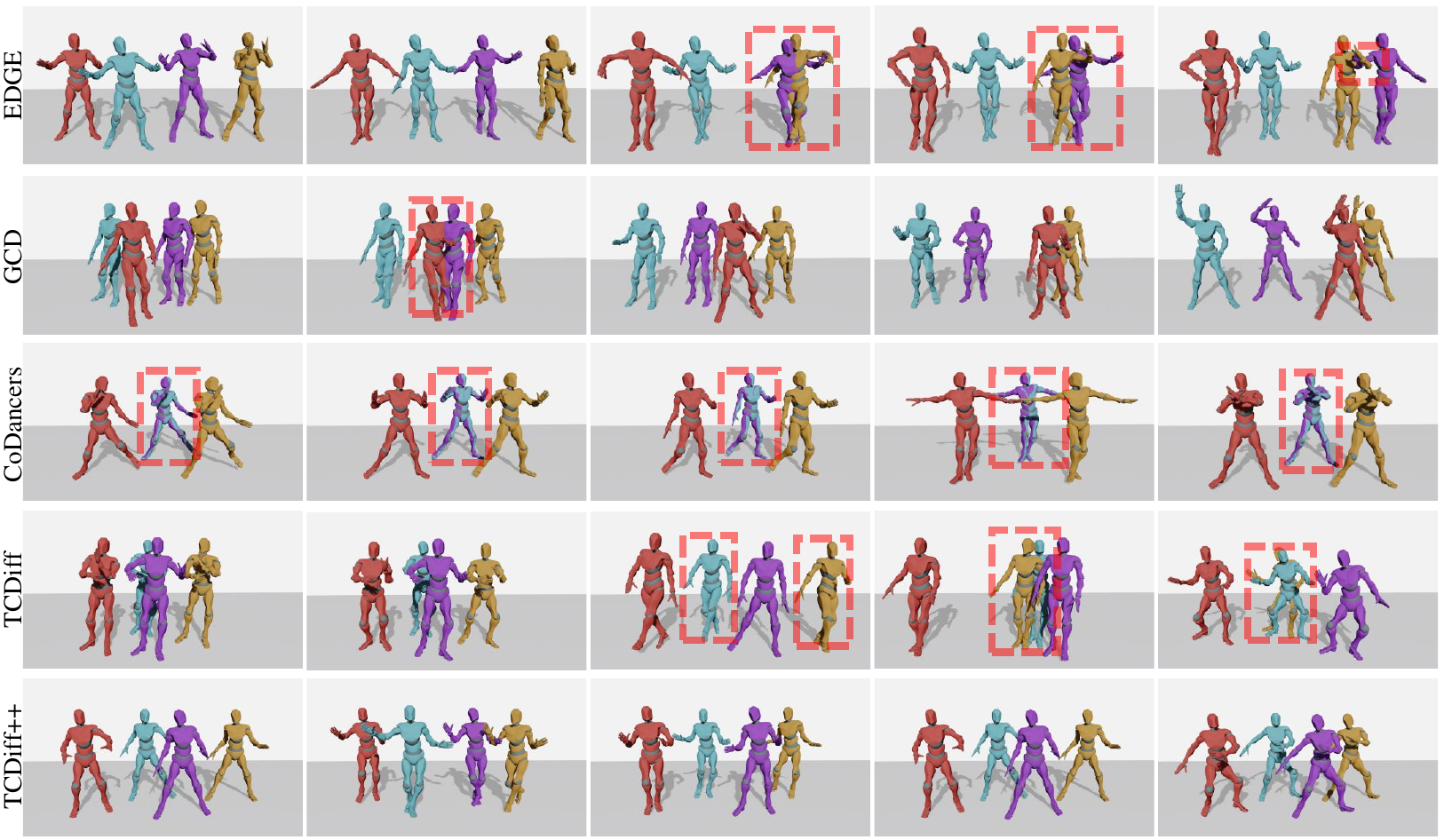}
  \caption{Visual comparison with Baselines. Baselines often cause collisions (highlighted in the \underline{red} box) during exchanges.
  }
  \label{fig:vis_comparison}
\end{figure*}

\section{Experiments}
\label{sec:Experiments}
\subsection{Experimental Settings}
\noindent \textbf{Implementation Details.}
\label{sec:ImplementationDetails}
Our model employs a diffusion-based architecture comprising the Group Dance Decoder (GDD) and the Footwork Adaptor (FA). We set the hidden size of all components in our framework to $d=512$.
In GDD, a linear layer first transforms the 151-dimensional input motion into a 512-dimensional hidden representation. This representation is then processed by the Fusion Projection, which consists of a three-layer MLP with ReLU activation. Next, the Sequence Decoder, composed of $M = 8$ stacked layers and equipped with eight attention heads, extracts sequential information from the data. Finally, an output linear layer maps the hidden features back to 151-dimensional raw motion data.
For footwork modulation, we construct the FA using $W = 3$ stacked Concat Squash Linear layers.
We set the weighting coefficients as follows: $\lambda_{\text{Sim}} = 0.636$, $\lambda_{\text{vel}} = 2.964$, $\lambda_{\text{FK}} = 0.646, \lambda_{\text{D}}= 100$, and $\lambda_{\text{con}} = 10.942$.
We used batch sizes of 72, 53, 35, and 23 per GPU for 2, 3, 4, and 5 dancers, respectively. We use Adam for optimization with a learning rate of $5 \times 10^{-5}$.

\textbf{Dataset.}
\label{sec:Dataset}
The open-source AIOZ-GDance dataset \citep{aioz} provides 16.7 hours of 3D multi-dancer motion that capture data for group dance performances, synchronized with musical accompaniment. Each video lasts 15 to 60 seconds and is decoded at 30 FPS. It features over 4,000 performers, ensuring diversity in motion and auditory content. Following the protocol in \citep{aioz}, videos are partitioned into training (80\%), validation (10\%), and test (10\%) sets.

\textbf{Compared methods.}
\label{sec:baselines} 
We evaluate TCDiff\texttt{++} against four publicly available group dance generation baselines: 
\begin{itemize}
    \item GCD \citep{gcd} leverages contrastive diffusion for controllable group dance generation, balancing diversity and coherence via user-guided adjustments.
    \item CoDancers \citep{codancers} decomposes group dance into individual solo dances, generating dancers sequentially and ultimately assembling them into a cohesive group performance.
    \item TCDiff \citep{tcdiff} is our previous version, a two-stage model that first generates dancers' trajectories and then their corresponding movements, enabling collision-free results in short-duration scenarios.
    \item We adapt EDGE \citep{edge} for further comparison, a leading single-dancer model, by training it on the AIOZ-GDance dataset to benchmark its generalization to group settings.
\end{itemize} 
To the best of our knowledge, these represent all the available group dance generation models capable of producing choreography for two or more dancers.
Our previous version, TCDiff, is the current state-of-the-art method. In this paper, we adopt an end-to-end design to enhance the consistency of footwork motion and improve its performance in long-term generation, achieving superior results.

\begin{figure*}[!t]
  \centering
  \includegraphics[width=0.98\linewidth]{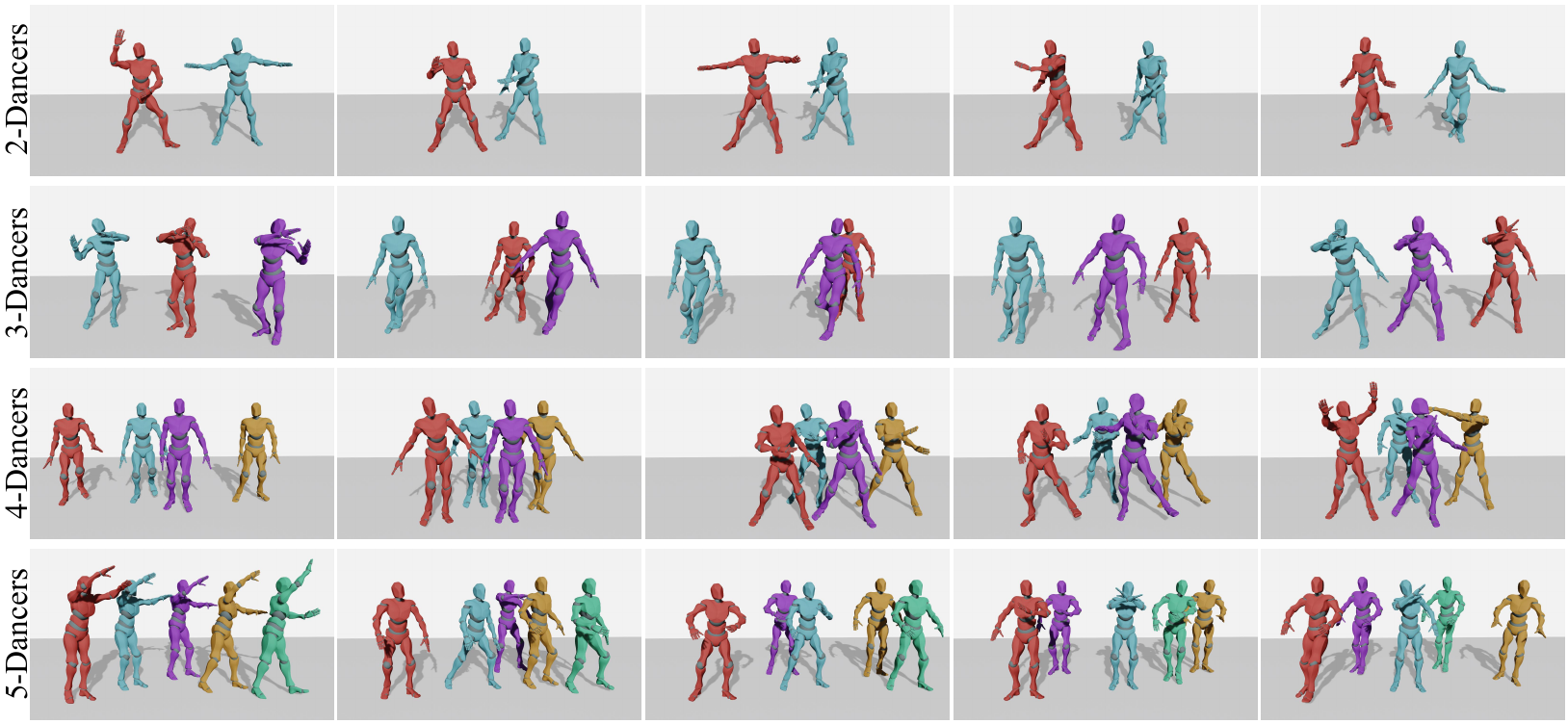}
  \caption{Results generated by TCDiff\texttt{++} with different numbers of dancers, demonstrating our adaptability to group sizes.
  }
  \label{fig:various_dancer}
\end{figure*}

\textbf{Metrics.}
\label{sec:Metrics}
We evaluate our model using metrics for both multi-dancer and single-dancer assessments. For multi-dancer evaluation~\citep{aioz}, (1) Group Motion Realism (GMR) measures group formation feature similarity using the Frechet Inception Distance (FID). (2) Group Motion Correlation (GMC) assesses coherence by calculating the cross-correlation between generated dancers. (3) The Trajectory Intersection Frequency (TIF) evaluates how often dancers collide during movement.
For single-dancer evaluation, (4) the Frechet Inception Distance (FID)~\citep{aist++, heusel2017gans} quantifies the similarity between individual dances and ground-truth dances. (5) Generation Diversity (Div)~\citep{aist++, dancerevolution} measures the variety of dance movements using kinematic features. (6) Motion-Music Consistency (MMC)~\citep{aist++} examines how well the generated dances synchronize with the music’s rhythm. (7) The Physical Foot Contact (PFC)~\citep{edge} score evaluates the physical plausibility of footwork by analyzing the relationship between the center of mass and foot velocity.

\subsection{Qualitative Visual Comparison}
\label{sec:visualcom}

\noindent \textbf{Baseline visual comparison. }
Figure \ref{fig:vis_comparison} illustrates the performance comparison between our model and the baselines.
The results indicate that the single-dancer model EDGE is affected by the dancer ambiguity phenomenon, leading to dancer collisions. This demonstrates the difficulty of directly applying single-dancer models to multi-dancer generation scenarios.
The group dance model GCD utilizes global attention mechanisms to capture dancer interactions. However, it neglects positional differences, resulting in multi-dancer collisions.
CoDancers produces unreasonable initial positions due to incomplete group information, resulting in severe overlaps.
TCDiff addresses the trajectory overlap issue caused by dancer ambiguity through a two-stage generation approach. However, this two-stage process decouples trajectory modeling from body modeling, resulting in suboptimal performance. This decoupling causes the dancer's coordinates to gradually converge, ultimately leading to collisions.
In contrast, TCDiff\texttt{++} employs an end-to-end generation approach, leveraging the design of internal modules to enhance the distinction between spatial features, thus resolving dancer ambiguity while yielding more coherent position-body movements.

\begin{table}[tb]
  \vspace{-8pt}
  \caption{Quantitative comparison with the baselines.}
  \setlength{\belowcaptionskip}{-1pt}
  \centering
  \setlength{\tabcolsep}{0.8pt} 
  \begin{tabular}{lcccccccc}
    \toprule
    \multirow{2}{*}{Method} & \multicolumn{3}{c}{Group-dance Metric} & \multicolumn{4}{c}{Single-dance Metric} \\
    \cmidrule(lr){2-4} 
    \cmidrule(lr){5-8}
    & GMR$\downarrow$ & GMC$\uparrow$ & TIF$\downarrow$ & FID$\downarrow$ & Div$\uparrow$ & MMC$\uparrow$ & PFC$\downarrow$  \\
    \midrule
    EDGE & 63.35 & 61.72 & 0.36 & 31.40 & 9.57 & 0.26 & 2.63 \\
    GCD & 31.47 & 80.97 & 0.17 & 31.16 & 10.87 & \textbf{0.26} & 2.53  \\
    CoDancers & 26.10 & 74.05 & \textbf{0.10} & \underline{23.98} & 9.48 & 0.25 & 3.26 \\
    TCDiff & \underline{13.86} & \underline{81.98} & \underline{0.13} & 37.47 & \underline{15.10} & 0.25 & \textbf{0.51} \\
    \midrule
    \textbf{TCDiff\texttt{++}} & \textbf{10.66} & \textbf{82.00} & \textbf{0.10} & \textbf{21.69} & \textbf{18.48} & \underline{0.25} & \underline{1.38} \\
    \bottomrule
  \end{tabular}
  \vspace{-2pt}
  \footnotetext{$\uparrow$ means higher is better, $\downarrow$ means lower is better. The best results are highlighted in bold, the second best results are underlined.}
  \label{tab:baselines}
\end{table}

\noindent \textbf{Varying Number Analysis. }
Figure~\ref{fig:various_dancer} demonstrates the capability of TCDiff\texttt{++} to generate high-quality dance motions across a range of group sizes. Our model not only preserves the realism and smoothness of individual motions but also effectively captures group-level dynamics and interactions. In the 5-dancer case, it produces well-balanced and visually harmonious formations. In smaller groups, such as 3 dancers, it generates meaningful interactions like positional swaps, while in the 2- and 4-dancer settings, it exhibits rich motion diversity and coordination patterns. These results highlight the model’s adaptability to varying group sizes and its ability to synthesize coherent choreography that reflects both individual expressiveness and collective structure.
More visualization examples can be found on our \href{https://da1yuqin.github.io/TCDiffpp.website/}{page}.

\begin{figure*}[!t]
  \centering
  \includegraphics[width=0.97\linewidth]{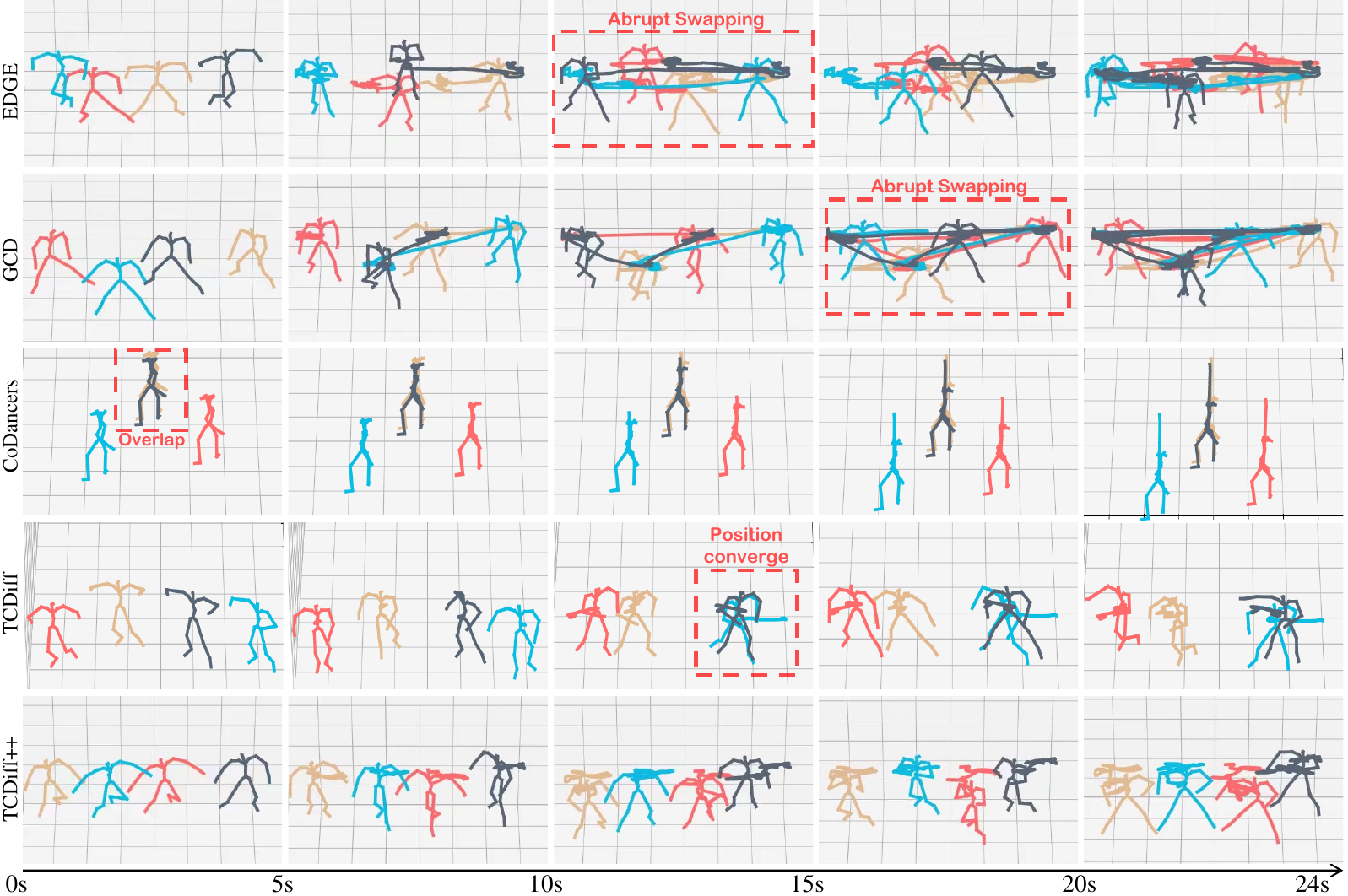}
  \caption{
  Baseline comparison with positional trajectory visualization in a long-duration setting.
  }
  \label{fig:longvis}
\end{figure*}

\noindent \textbf{Long-duration scenario comparison. }
We present a visual comparison of the baseline's long-duration generation (720 frames) in Figure~\ref{fig:longvis}, to better illustrate the superiority of TCDiff\texttt{++} in this scenario. 
To better illustrate the dancer formations, we visualize not only the dancers' pose skeletons but also their positional trajectories, providing a more comprehensive representation of the dynamic formation transitions.
Due to the influence of dancer swap phenomena, methods that adopt naive long diffusion sampling (EDGE, GCD) produce severe abrupt swapping during long-duration generation. This is because, during the sampling phase, the lack of spatial information leads to inconsistent dancer position generation across different epochs, causing abrupt swapping when merging the generated segments.
Since CoDancers generates one dancer at a time, it lacks inter-dancer information, such as positions,  
As a result, CoDancers produces unreasonable initial positions due to incomplete group information, resulting in severe overlaps.
Moreover, the dancers tend to perform nearly identical movements with minimal interaction, losing the collaborative dynamics that distinguish group dance from solo performances.
TCDiff generates dancer positions without considering their movements, which makes it prone to introducing erroneous position estimates due to uncertainty. 
These errors accumulate significantly during long-duration group dance generation, causing the dancer coordinates to gradually converge, ultimately resulting in collisions and rendering TCDiff unsuitable for extended sequences.
In contrast, the end-to-end TCDiff\texttt{++} takes into account previously generated results during expansion, improving position consistency across epochs. This effectively reduces the occurrence of abrupt swapping phenomena.

\subsection{Quantitative Comparison}
\label{sec:quancomparison}
\begin{figure}[!t]
  \centering
  \includegraphics[width=0.98\linewidth]{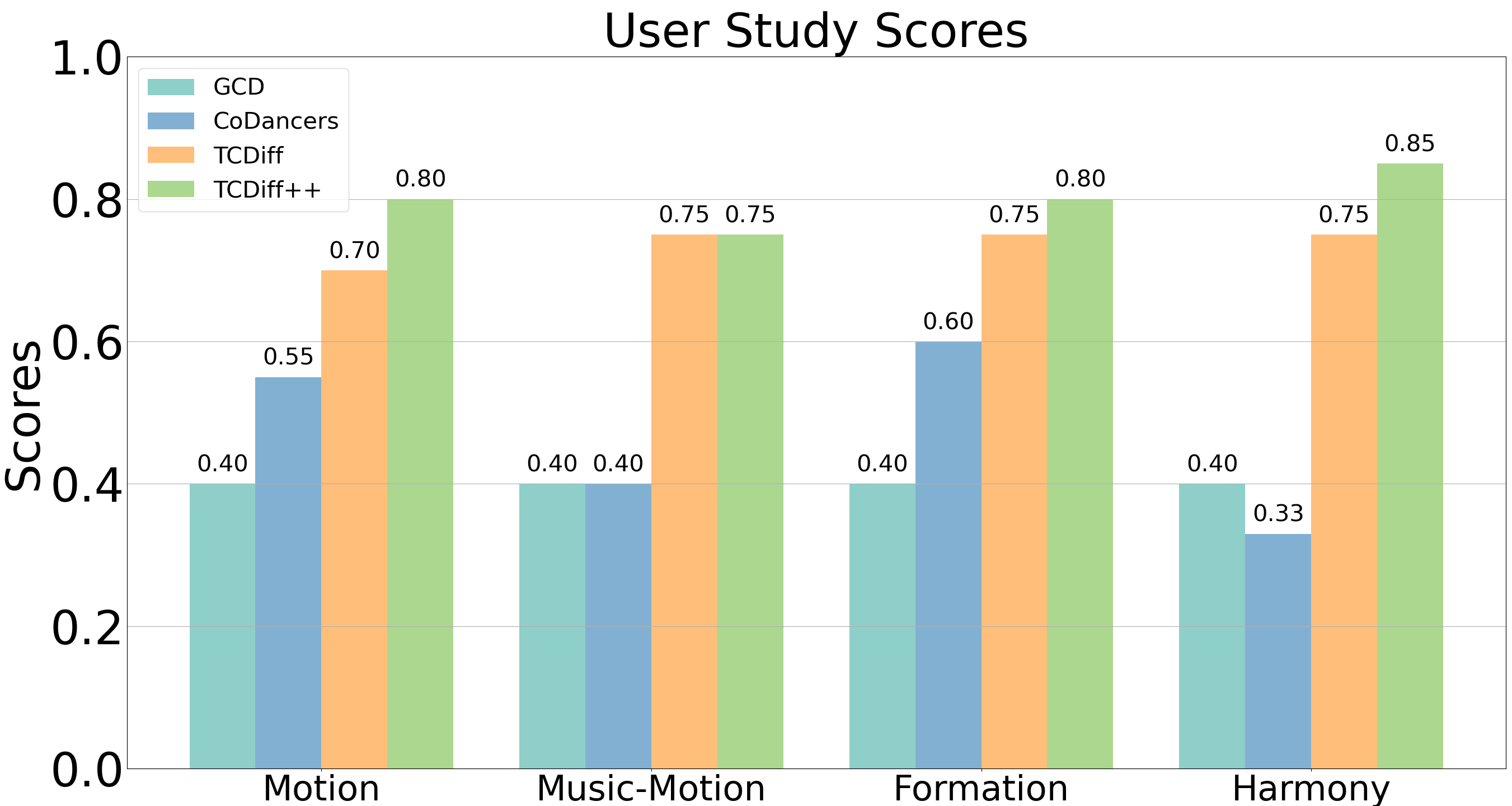}
  \caption{User Study Results. Our model received higher user ratings, demonstrating its superior aesthetic appeal compared to other methods.
  }
  \label{fig:userstudy}
\end{figure}

\noindent \textbf{Baseline comparison. }
Tables \ref{tab:baselines} and \ref{tab:dancernummetrics} compare our model's performance with baseline methods. 
The single-dancer model EDGE struggles in multi-person scenarios, exhibiting severe foot sliding and frequent multi-dancer collisions (high TIF and PFC) due to dancer ambiguity.
Similarly, GCD excessively focuses on inter-dancer interaction while neglecting the modeling of coordinate differences between dancers, making it susceptible to dancer ambiguity and resulting in severe foot sliding (high PFC).
CoDancers~\citep{codancers} reduces ambiguity (low TIF) but compromises inter-dancer correlations (low GMC) and formation integrity, resulting in discordant group formations, as shown in our user study (Figure \ref{fig:userstudy}).
Additionally, CoDancers overlooks the relationship between footwork and motion, limiting its ability to align footwork actions with positional changes, which hinders the generation of accurate footwork (low PFC). 
By decoupling dancer coordinates and movements into two stages, our previous version, TCDiff, mitigates ambiguity, improves coordination, and enhances formation quality. 
However, this two-stage generation process decouples position trajectory from body modeling, resulting in disjointed actions and displacements, which compromise individual fidelity (FID) and lead to suboptimal results.
In contrast, TCDiff\texttt{++} adopts an end-to-end architecture to mitigate disjoint generation, leveraging its internal module design to resolve dancer ambiguity. This enables more coherent positioning and body movements, leading to consistent superiority in group dance metrics and exceptional performance in Div and FID for single-dance metrics.

\begin{footnotesize}     
\begin{table}[!t]
  \centering
  \vspace{3pt}
  \caption{Quantitative comparison with the baselines for generating 2-5 dancers.}
  \setlength{\belowcaptionskip}{-1pt}
  \setlength{\tabcolsep}{2pt} 
  \begin{tabular}{lccccccccc}
    \toprule
     Method & \# & GMR$\downarrow$ & GMC$\uparrow$ & TIF$\downarrow$ &  FID$\downarrow$ & Div$\uparrow$ & MMC$\uparrow$ \\
    \midrule
    \multirow{4}{*}{GCD} & 2 & 34.09 & 80.26 & 0.167 & 32.62 & 10.41 & 0.266 \\
    & 3 & 36.25 & 79.93 & 0.184 & 33.94& 10.02 & 0.266 \\
    & 4 & 36.28 & 81.82 & 0.125 & 35.89 & 9.87 & 0.251 \\
    & 5 & 38.43 & 81.44 & 0.168 & 35.08 & 9.92 & 0.264 \\
    \midrule
    \multirow{4}{*}{CoDancers} & 2 & 24.53 & 72.88 & 0.080 & 26.31 & 9.01 & 0.251 \\
    & 3 & 27.23 & 74.34 & 0.084 & 24.85 & 9.15 & 0.254 \\
    & 4 & 26.44 & 75.34 & 0.097 & 25.76 & 9.43 & 0.258 \\
    & 5 & 26.34 & 74.22 & 0.113 & 25.45 & 9.77 & 0.253 \\
    \midrule
    \multirow{4}{*}{TCDiff} & 2 & 15.77 & 81.92 & 0.121 & 41.26 & 16.20 & 0.263 \\
    & 3 & 10.97 & 81.51 & 0.123 & 48.00 & 19.28 & 0.253 \\
    & 4 & 13.44 & 81.70 & 0.149 & 23.32 & 10.89 & 0.253 \\
    & 5 & 15.36 & 82.77 & 0.109 & 37.31 & 14.01 & 0.236 \\
    \midrule
    \multirow{4}{*}{\textbf{TCDiff\texttt{++}}} & 2 & 12.07 & 82.73 & 0.058 & 18.75 & 14.87 & 0.249 \\
    & 3 & 9.75 & 81.23 & 0.098 & 23.75 & 22.04 & 0.251 \\
    & 4 & 10.02 & 82.70 & 0.143 & 20.37 & 15.15 & 0.252 \\
    & 5 & 10.77 & 81.37 & 0.106 & 23.91 & 21.87 & 0.266 \\
    \bottomrule
  \end{tabular}
  \vspace{-2pt}
  \footnotetext{$\uparrow$ means higher is better, $\downarrow$ means lower is better.}
  \label{tab:dancernummetrics}
\end{table}
\end{footnotesize}

\begin{figure*}[!t]
  \centering
  \includegraphics[width=0.99\linewidth]{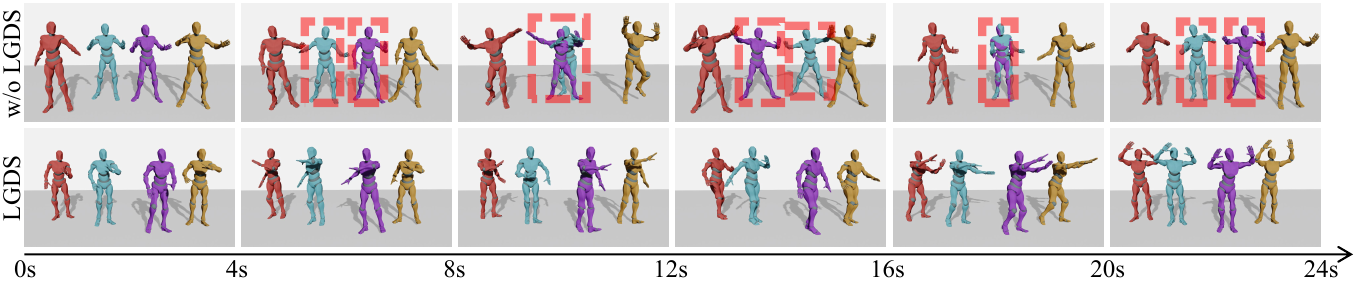}
  \caption{Comparison of group dance generation results with and without the LGDS module under long-duration conditions.
  LGDS effectively reduces abrupt dancer swapping in long-term generation, enhancing spatial consistency.
  }
  \label{fig:LGDS_ab}
\end{figure*}

\noindent \textbf{User study }
Our user study results are presented in Figure~\ref{fig:userstudy}.
The study involved participants primarily from a younger demographic (ages 18–60, mostly in their 20s and 30s), aligning with the target user group of our application scenario.
Regarding the experimental procedure, we anonymized the outputs generated by different methods and randomly mixed them before presenting them to participants. They were then asked to rate the outputs on a scale from 1 to 10 based on four criteria: motion realism, music-motion correlation, formation aesthetics, and dancer harmony. The final voting results were normalized to a $[0, 1]$ range for consistent comparison.  
To ensure fairness, none of the samples from the previous conference version were reused, ensuring that all outputs remained anonymous to users.
As a result, slight differences in scores may occur.
Nevertheless, the results indicate that the visual effects generated by our model are the most favored by users, as TCDiff\texttt{++} effectively reduces visual artifacts such as collisions and foot sliding, while producing more consistent footwork and harmonious group formations.

\begin{table}[tb]
  \centering
  \vspace{-8pt}
  \caption{
  Baseline comparison in a long-duration setting.
  }
  \setlength{\belowcaptionskip}{-1pt}
  \setlength{\tabcolsep}{0.8pt} 
  \begin{tabular}{lcccccccc}
    \toprule
    \multirow{2}{*}{Method} & \multicolumn{3}{c}{Group-dance Metric} & \multicolumn{4}{c}{Single-dance Metric} \\
    \cmidrule(lr){2-4} 
    \cmidrule(lr){5-8}
    & GMR$\downarrow$ & GMC$\uparrow$ & TIF$\downarrow$ & FID$\downarrow$ & Div$\uparrow$ & MMC$\uparrow$ & PFC$\downarrow$  \\
    \midrule
    EDGE & 67.24 & 57.65 & 0.38 & 35.40 & 7.97 & \textbf{0.24} & 3.87 \\
    GCD & 40.68 & 79.25 & 0.28 & 51.25 & 7.24 & 0.20 & 3.52  \\
    CoDancers & 35.20 & 70.53 & 0.15 & 43.62 & 5.48 & 0.22 & 4.23 \\
    TCDiff & 23.87 & 80.97 & 0.17 & 50.47 & 14.30 & 0.23 & 1.87 \\
    \midrule
    \textbf{TCDiff\texttt{++}} & \textbf{14.67} & \textbf{81.64} & \textbf{0.15} & \textbf{20.37} & \textbf{16.19} & 0.23 & \textbf{1.53} \\
    \bottomrule
  \end{tabular}
  \footnotetext{$\uparrow$ means higher is better, $\downarrow$ means lower is better. The best results are highlighted in bold.}
  \label{tab:longbaselines}
\end{table}

\noindent \textbf{Evaluating long-duration scenarios. }
To assess long-duration performance, we extend the test to 720 frames and evaluate key metrics (Table~\ref{tab:longbaselines}).
It can be observed that all models exhibit varying degrees of performance degradation in long-term generation. This includes issues such as frozen frames (resulting in lower MMC) and abrupt swaps (leading to performance declines in PFC and TIF).
Due to dancer swap phenomena, naive long diffusion sampling (EDGE, GCD) causes severe abrupt swapping in long-duration generation, as the absence of spatial information disrupts position consistency across epochs.
Among them, due to primarily focuses on individual dancers, EDGE maintains more stable motion quality (MMC). However, this comes at the cost of overlooking global feature, leading to more severe dancer collisions (higher TIF) and poor group dance performance.
Both CoDancers and the first stage of TCDiff adopt an autoregressive approach, where TCDiff generates trajectory coordinates, while CoDancers synthesize complete dance motions.
However, CoDancers solely focuses on individual-level information while neglecting group-level features, leading to degraded performance on group-dance metrics.
TCDiff estimates dancer positions without explicitly modeling movements, making it prone to errors from movement uncertainty. These accumulate over time, causing disjointed motions and positional inconsistencies, limiting its suitability for long-duration generation.
In contrast, TCDiff\texttt{++} employs an end-to-end design to improve the coherence between positions and body movements, achieving the best long-duration performance. It also integrates past results to maintain positional consistency and mitigate abrupt swapping.

\subsection{Ablation Study}
\label{sec:ablationstudy}

\begin{table*}[h]
  \caption{Ablation study evaluating the incremental contributions of each component in TCDiff\texttt{++}. The best results are highlighted in bold, the second best results are underlined}
  \label{tab:incremental_ablation}
  \centering
  \resizebox{0.95\textwidth}{!}{%
  \begin{tabular}{lcccccccc}
    \toprule
    \multirow{2}{*}{Method} & \multicolumn{3}{c}{Group-dance Metric} & \multicolumn{4}{c}{Single-dance Metric} \\
    \cmidrule(lr){2-4} 
    \cmidrule(lr){5-8}
    & GMR$\downarrow$ & GMC$\uparrow$ & TIF$\downarrow$ & FID$\downarrow$ & Div$\uparrow$ & MMC$\uparrow$ & PFC$\downarrow$ \\
    \midrule
    Base & 67.24 & 57.65 & 0.38 & 55.40 & \textbf{19.97} & 0.23 & 3.87 \\
    +FP & 50.12 & 77.80 & 0.31 & 44.31 & \underline{16.90} & 0.20 & 3.01 \\
    +FP+DPE & 44.38 & 78.22 & 0.28 & 40.75 & 16.82 & 0.22 & 3.08 \\
    +FP+DPE+$\mathcal{L}_D$ & 25.76 & 81.01 & 0.19 & 30.97 & 12.45 & 0.21 & 2.54 \\
    +FP+DPE+$\mathcal{L}_D$+SM & 23.93 & 81.45 & 0.17 & 27.61 & 12.88 & 0.21 & 2.60 \\
    +FP+DPE+$\mathcal{L}_D$+SM+LGDS & \underline{23.52} & 81.48 & 0.16 & 27.50 & 12.90 & 0.22 & 2.48 \\
    +FP+DPE+$\mathcal{L}_D$+SM+LGDS+FA & 23.95 & \underline{81.52} & \underline{0.16} & \underline{24.75} & 14.88 & \textbf{0.24} & \textbf{1.49} \\
    +FP+DPE+$\mathcal{L}_D$+SM+LGDS+FA+SD (Ours) & \textbf{14.67} & \textbf{81.64} & \textbf{0.15} & \textbf{20.37} & 16.19 & \underline{0.23} & \underline{1.53} \\
    \bottomrule
  \end{tabular}
  }
  \vspace{-2pt}
\end{table*}

\begin{figure}[!t]
  \centering
  \includegraphics[width=0.99\linewidth]{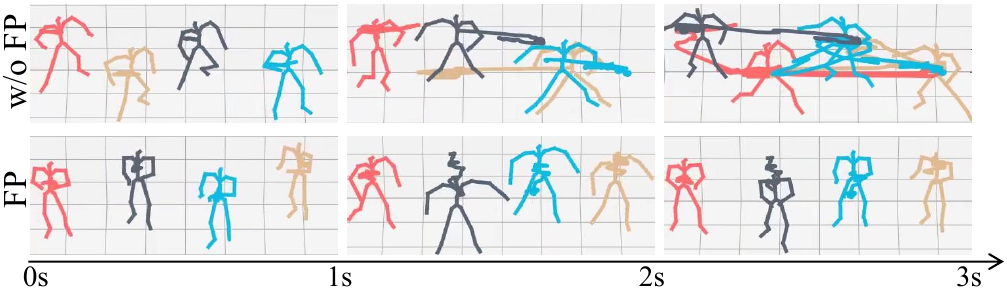}
  \vspace{-5mm}
  \caption{Comparison with and without the FP module in our end-to-end TCDiff\texttt{++}. The skeletal motion visualization demonstrates that the FP module effectively mitigates overlap and abrupt swapping phenomena.}
  \label{fig:fusion_projection_ab}
\end{figure}

We conduct a step-by-step ablation to evaluate the incremental contributions of each module. We provide visual comparisons in our demo video, available on our \href{https://da1yuqin.github.io/TCDiffpp.website/}{project page}, for a more intuitive understanding of the improvements. 
All modules bring clear and cumulative improvements, both quantitatively and visually. Our ablation study not only presents clear quantitative improvements but also highlights how each module addresses key visual challenges in long-duration group dance generation. Detailed results are shown in Table~\ref{tab:incremental_ablation}. Note that the high diversity score of the Base model is largely due to irregular abrupt swaps, which do not indicate good visual quality. These severe artifacts are significantly reduced after introducing $\mathcal{L}_D$. Below is a concise analysis of each component:

\textbf{Baseline Model:}  
The baseline model~\citep{edge} exhibits significant motion instability, with frequent abrupt position swaps and severe jitter in long sequences. This results in high GMR (67.24), poor group coherence, with a GMC score of 57.65, and high TIF (0.38), indicating that dancers often collide or lose spatial consistency. Note that the high diversity score is largely due to irregular abrupt swaps, which do not indicate good visual quality.

\textbf{Impact of Fusion Projection (FP):}  
\label{sec:fpab}
FP enhances the model's ability to distinguish between dancers by projecting their features into a shared high-dimensional space. This leads to a clear improvement in group motion quality, reflected in a reduction of GMR from 67.24 to 50.12 and an increase in GMC from 57.65 to 77.80. Additionally, it helps slightly reduce foot collisions, as indicated by the PFC decrease from 3.87 to 3.01.

\textbf{Impact of Dancer Positioning Embedding (DPE):}  
\label{sec:dpeab}
The introduction of DPE stabilizes the dancers' trajectories by embedding both temporal and identity information. This reduces jitter, improving the overall visual stability, as shown by further reductions in GMR (50.12 $\rightarrow$ 44.38) and TIF (0.31 $\rightarrow$ 0.28). However, PFC slightly increases from 3.01 to 3.08, suggesting that while overall group structure improves, foot sliding issues persist without additional motion constraints.

\textbf{Impact of Distance-Consistency Loss ($\mathcal{L}_D$):}  
\label{sec:dlab}
$\mathcal{L}_D$ imposes global spatial constraints that maintain consistent inter-dancer spacing. This significantly reduces collisions, driving GMR down from 44.38 to 25.76 and TIF from 0.28 to 0.19. It also lowers PFC (3.08 $\rightarrow$ 2.54), indicating less foot sliding. Nevertheless, due to the probabilistic nature of diffusion, occasional abrupt swaps can still occur in long sequences.

\textbf{Impact of Swap Mode Embedding (SM):}  
\label{sec:smab}
SM encodes dancer positional information at segment boundaries, effectively reducing confusion in spatial transitions. This results in improved formation and decreased TIF (0.19 $\rightarrow$ 0.17). GMC also improves slightly (81.01 $\rightarrow$ 81.45), confirming better temporal coordination during swaps.

\textbf{Impact of Long Group Diffusion Sampling (LGDS):}  
\label{sec:lgdsab}
LGDS enhances long-term temporal consistency by conditioning each segment on previously generated frames. This stabilizes continuity, particularly in long sequences. The main visual improvement lies in the further reduction of severe abrupt swaps. GMR slightly improves (23.93 $\rightarrow$ 23.52), and TIF decreases to 0.16, indicating fewer inter-dancer collisions. PFC also drops from 2.60 to 2.48, reinforcing better lower-body coherence.

\textbf{Impact of Footwork Adaptor (FA):}  
\label{sec:faab}
FA focuses on refining lower-body movement by aligning foot trajectories with position shifts. This leads to a significant drop in PFC (2.48 $\rightarrow$ 1.49), confirming reduced foot sliding and improving the physical realism of foot placement. MMC improves from 0.22 to 0.24, indicating better alignment with music rhythms, while other metrics (e.g., FID drops from 27.50 to 24.75) also suggest enhanced visual quality.

\textbf{Impact of Sequence Decoder (SD):}  
\label{sec:sdab}
The SD replaces the Transformer with a State Space Model, enabling more efficient and coherent long-sequence modeling. This results in the largest gains: GMR drops significantly from 23.95 to 14.67, FID improves from 24.75 to 20.37, and Div increases from 14.88 to 16.19. Despite a slight increase in PFC (1.49 $\rightarrow$ 1.53), the generated dances exhibit more fluid, expressive, and temporally consistent motion.

In conclusion, 
each component in our model addresses specific visual issues, from jitter and foot sliding to positional swaps and long-duration coherence. FP and DPE improve spatial awareness; $\mathcal{L}_D$, SM, and LGDS enhance group formation and reduce swapping; FA reduces foot sliding and enhances physical realism; and SD boosts long-term coherence and expressiveness. These improvements collectively yield the best performance across all group- and single-dancer metrics, validating the design of our modular architecture.

\section{Limitation}
\noindent 
Due to the early stage of group dance research, our work still has limitations. 
First, our model lacks support for multimodal or user-controllable inputs. It focuses solely on the basic cross-modal generation task (e.g., generating dance from music), without incorporating any additional control signals. While this provides a necessary foundation, real-world applications often require richer forms of guidance, such as textual descriptions, motion keyframes, or genre-specific conditioning, to better reflect user intent. Given the ongoing challenges in visual quality and temporal consistency, we prioritize solving the core generation problem in this work. We leave multimodal and interactive control for future exploration, which we believe is essential for practical deployment, especially in complex group dance scenarios.
Second, our model shows limited capability in learning effective swap patterns. This arises partly because modeling swap actions is still in its infancy and, more crucially, because existing datasets lack sufficient samples and explicit annotations of swap behaviors. As a result, performance on swap-related generation remains constrained. We believe future work should improve both model architectures and data resources to better capture and generate swap behaviors in group choreography.

\section{Conclusion}
 \noindent This paper presents TCDiff\texttt{++}, an innovative framework for generating harmonious group dance through a music-driven, end-to-end approach. We address key challenges in group dance generation, including multi-dancer collisions, single-dancer foot sliding, and long-duration generation. By incorporating dancer positioning embeddings and distance-consistency loss, we effectively reduce collisions and maintain relative positioning. The introduction of swap mode embedding and the Footwork Adaptor minimizes foot sliding, enhancing the quality of individual dancer motions. Furthermore, our long group diffusion sampling strategy, along with the Sequence Decoder layer, ensures smooth and consistent long-duration generation. Experimental results demonstrate that TCDiff\texttt{++} outperforms existing methods, offering significant improvements in handling long-duration group dance generation while preserving both individual and collective motion quality.

\section*{Declarations}
\bmhead{Competing interests}The authors declare that they have no known competing financial interests or personal relationships that could have appeared to influence the work reported in this paper.
\bmhead{Data availability}This work does not propose any new dataset. The dataset~\citep{aioz} that support the findings of this study is
openly available at the URL: \href{https://github.com/aioz-ai/AIOZ-GDANCE}{https://github.com/aioz-ai/AIOZ-GDANCE}.
\bmhead{Acknowledgments}
This work was supported by the National Science Fund of China under Grant Nos. U24A20330, 62361166670 and 62072242.

\bibliography{ijcv25}

\end{document}